\documentclass{kluwer}

\usepackage{dcolumn}   
\usepackage{bm}        
\usepackage{latexsym}
\usepackage{amsfonts}
\usepackage{amssymb}
\usepackage{graphicx}

\def\la{\;\raise0.3ex\hbox{$<$\kern-0.75em\raise-1.1ex\hbox{$\sim$}}\;}
\def\ga{\;\raise0.3ex\hbox{$>$\kern-0.75em\raise-1.1ex\hbox{$\sim$}}\;}

\def\a{\alpha}
\def\adef{\,\alpha=e^2/\hbar c\,}
\def\mudef{\,\mu=m_p/m_e\,}
\def\h2{H$_2$}

\begin{document}
\begin{article}

\begin{opening}
\title{Does the proton-to-electron mass ratio $\mudef$ vary in the course of cosmological evolution? }
\author{A. \surname{Ivanchik}}
\institute{Ioffe Physical-Technical Institute, St.-Petersburg,
Russia}
\author{P. \surname{Petitjean}}
\author{E. \surname{Rodriguez}}
\institute{Institut d'Astrophysique de Paris -- CNRS, Paris,
France}
\author{D. \surname{Varshalovich}}
\institute{Ioffe Physical-Technical Institute, St.-Petersburg,
Russia}

\runningtitle{Proton-to-electron mass ratio $\mudef$}

\runningauthor{A. Ivanchik et al.}

\date{September 4, 2002}

\begin{abstract}
The possible cosmological variation of the proton-to-electron mass
ratio $\mudef$ was estimated by measuring the \h2 wavelengths in
the high-resolution spectrum of the quasar Q~0347-382. Our
analysis yielded an estimate for the possible deviation of
$\,\mu\,$ value in the past, 10~Gyr ago: for the unweighted value
$\Delta \mu / \mu = (3.0\pm2.4)\times10^{-5}$; for the weighted
value
\[
\Delta \mu / \mu = (5.02\pm1.82)\times10^{-5}\, .
\]
Since the significance of the both results does not exceed
3$\sigma$, further observations are needed to increase the
statistical significance. In any case, this result may be
considered as the most stringent estimate on an upper limit of a
possible variation of $\mu$ (95\% C.L.):
\[
|\Delta \mu / \mu| < 8\times 10^{-5}\, .
\]
This value serves as an effective tool for selection of models
determining a relation between possible cosmological deviations of
the fine-structure constant $\alpha$ and the elementary particle
masses (m$_p$, m$_e$, etc.).
\end{abstract}

\keywords{cosmology, quasar absorption spectra, fundamental
physical constants}

\end{opening}

\section*{Introduction}
Contemporary theories of fundamental interactions
(Strings/M-theory,  and others) predict variation of fundamental
physical constants in the course of the Universe evolution.

First of all the theories predict variation of the coupling
constants with increasing of energy transfer in particle
interaction (so-called, ``running constants''). This effect has
been proved in high-energy experiments with accelerators. For
example, the fine-structure constant $\adef$ equals $1/137.036$ at
low energies $(E\rightarrow 0)$, but it increases up-to
$1/128.896$ for energy $E\sim 90$~GeV. Such ``running'' of the
constants has to be taken into account for consideration of very
early Universe.

There is another prediction of the current theories related to the
low-energy limits of the constants. These limits are allowed to
vary during the cosmological evolution and even have different
values in different points of the space-time. There are several
reasons for such variations. Multidimensional theories
(Kaluza-Klein models, ``p-brane'' models, and others) predict
variations of fundamental physical constants as a direct result of
the cosmological evolution of the extra-dimensional subspace. In
several theories (e.g. Superstrings), the variations of the
constants result from the cosmological evolution of the vacuum
state (a vacuum condensate of some scalar field or
``Quintessence'').

Here we check a possible cosmological deviation $\,\Delta \mu =
\mu - \mu_0$, where $\,\mu\,$ is the proton-to-electron mass ratio
at the epoch $z=3.0249$ and $\mu_0$ is the present one. Searching
of $\,\mu$-variation becomes especially important after possible
finding of $\,\alpha$-variation \cite{Webb} because different
theoretical models of the fundamental physical interactions
predict different variations of their values and different
relations between cosmological deviations of the constants
($\alpha$, $\mu$, $\Lambda_{QCD}$, and others
\cite{Langacker,Calmet}). Note that the method used for checking
of $\alpha$-variation by Webb et al. (2001) is based on a
simultaneous measurements of wavelengths of a large number of
transitions for various species. It is significantly reduced
statistical errors. However, it was more difficult to estimate the
systematic errors \cite{Murphy} than in the method used previously
where the structure of lines of each species was measured
separately \cite{IAV}. In any case, this intriguing result
\cite{Webb} must be checked independently by using different
methods and by probing variations of other fundamental constants.

Experimental detection of such variations of the constants would
be a great step forward in our understanding of Nature.

\section*{Proton-to-electron mass ratio $\mudef$}

At the present the proton-to-electron mass ratio has been measured
with a relative accuracy of $2\times 10^{-9}$ and equals $\mu_0 =
1836.1526670(39)$ \cite{Mohr}. Laboratory metrological
measurements rule out considerable variation of $\mu$ on a short
time scale but do not exclude its changes over the cosmological
scale, $\sim 10^{10}$ years. Moreover, one can not reject a
possibility of $\mu$ value (as well as other constants) be differ
in wide separated regions of the Universe.
It can be proved or disproved only by
means astrophysical observations of distant extragalactic objects.

Proper measurements of wavelengths of absorption lines in spectra
of high-redshift quasars provide us a tool for direct checking of
possible deviation of physical constants (e.g. $\mu$ and $\a$) at
the epoch of the absorption-spectrum formation, i.e. $\sim
10-13$~Gyr ago.

Some estimations of $\mu$-variations were performed by Potekhin et
al. (1999) and Levshakov et al. (2002a). The most stringent
estimate of possible cosmological variation of $\Delta \mu/\mu =
(5.7\pm3.8)\times 10^{-5}$ was obtained by Ivanchik et al. (2002).

\section*{Sensitivity Coefficients}
The method used here to determine possible $\mu$-variations was
proposed by Varshalovich and Levshakov (1993). It is based on the
fact that wavelengths of electron-vibro-rotational lines depend on
the reduced mass of the molecule, with the dependence being
different for different transitions. It enables to distinguish the
cosmological redshift of a line from the shift caused by a
possible variation of $\mu$.

Thus, the measured wavelength $\lambda_i$ of a line formed in the
absorption system at the redshift $z_{abs}$ can be written as
\begin{equation}
\lambda_i=\lambda_i^0(1+z_{abs})(1+K_i \Delta\mu/\mu) \, ,
\end{equation}
where $\lambda_i^0$ is the laboratory (vacuum) wavelength of the
transition, and $K_i= \mbox{d} \ln \lambda_i^0/ \mbox{d} \ln \mu$
is the sensitivity coefficient calculated for the Lyman and Werner
bands of molecular hydrogen in works \cite{VL}, \cite{VP}. This
expression can be represented in terms of the redshift $z_i \equiv
\lambda_i/\lambda_i^0-1$ as
\begin{equation}
z_i=z_{abs}+b \!\cdot \! \! K_i \, , \label{korr}
\end{equation}
where $b=(1+z_{abs})\Delta\mu/\mu$. In reality, $z_i$ is measured
with some uncertainty which is caused by statistical errors of the
astronomical measurements $\lambda_i$ and by errors of the
laboratory measurements of $\lambda_i^0$. Nevertheless, if $\Delta
\mu / \mu$ is nonzero, there must be a correlation between $z_i$
and $K_i$. Thus, a linear regression analysis of these quantities
yields $z_{abs}$ and $b$ (as well as its statistical
significance), consequently $\Delta\mu/\mu$.

\section*{Observations and Results}
High-resolution (FWHM $\approx$ 6~km/s) spectrum of the quasar
Q~0347-382 obtained with the 8.2-m VLT/UVES KUEYEN telescope (ESO)
were used to probe the possible variation of $\mu$. The observing
time was nine hours that led to S/N $\sim20-40$. We analyzed the
H$_2$ absorption system at $z_{abs}=3.0249$ in the spectrum. This
H$_2$ system was first detected by Levshakov et al. (2002b). More
than 80 lines of molecular hydrogen can be identified in the
wavelength range 3600-4600 \AA. For our analysis, we carefully
selected lines that satisfied the following conditions: (i)
isolated, and (ii) unblended. In this system, only 15 lines
satisfy these conditions. Parameters of the lines are given in
Table~I.
  \begin{table}[t]
   \small
    \caption{\small H$_2$ lines of absorption system at $z_{abs}=3.0249$ in Q~0347-382 spectrum}
       \begin{tabular}{rrrcr}
         \hline
         \noalign{\smallskip}
          Lines~~~~~ & K$_i$~~~ & $\lambda_i$, \AA~~~ & $\sigma(\lambda_i)$, \AA & $\lambda_i^0$, \AA~~~ \\
         \hline
         \noalign{\smallskip}
\mbox{W~~3-0~Q(1)} &  0.02176 & 3813.280 & 0.003 &  947.4218 \\
\mbox{W~~3-0~P(3)} &  0.01724 & 3830.381 & 0.004 &  951.6722 \\
\mbox{W~~2-0~Q(1)} &  0.01423 & 3888.436 & 0.003 &  966.0937 \\
\mbox{L~12-0~R(3)} &  0.04386 & 3894.792 & 0.009 &  967.6752 \\
\mbox{W~~2-0~Q(1)} &  0.01120 & 3900.322 & 0.002 &  969.0486 \\
\mbox{L~10-0~R(1)} &  0.04126 & 3952.749 & 0.003 &  982.0728 \\
\mbox{L~10-0~P(1)} &  0.04053 & 3955.814 & 0.002 &  982.8340 \\
\mbox{L~~9-0~P(1)} &  0.03719 & 3995.961 & 0.002 &  992.8087 \\
\mbox{W~~0-0~Q(2)} & -0.00686 & 4068.920 & 0.003 & 1010.9380 \\
\mbox{L~~6-0~R(3)} &  0.02262 & 4141.561 & 0.003 & 1028.9832 \\
\mbox{L~~4-0~R(3)} &  0.01304 & 4242.158 & 0.002 & 1053.9770 \\
\mbox{L~~4-0~P(3)} &  0.01071 & 4252.194 & 0.002 & 1056.4727 \\
\mbox{L~~3-0~P(1)} &  0.01026 & 4284.930 & 0.002 & 1064.6056 \\
\mbox{L~~3-0~R(3)} &  0.00758 & 4296.488 & 0.005 & 1067.4738 \\
\mbox{L~~2-0~P(3)} & -0.00098 & 4365.242 & 0.004 & 1084.5618 \\
\noalign{\smallskip} \hline
\end{tabular}
\end{table}
The average uncertainty of the determination of the line centers
is $\sigma(\lambda_i) \sim 3$~m\AA.

The result of the linear regression analysis of $z_i$-to-$K_i$ for
the \h2 lines presented in Table~I is shown on Fig.~1. It is
corresponding the following: $\Delta \mu / \mu =
(5.02\pm1.82)\times 10^{-5}$. This result was obtained taking into
account weighted coefficients for each line analyzed. The
unweighted value is $\Delta \mu / \mu = (3.0\pm2.4)\times
10^{-5}$.

The laboratory wavelengths were taken from works \cite{Abgr,Ron}.
The statistical uncertainties of the laboratory wavelengths are
about $1.5\,$m\AA~ corresponding to an error of $\Delta \mu / \mu$
about $2\times 10^{-5}$ that is in agreement with the errors found
from the regression.

\begin{figure}[h]
 \centering
   \includegraphics[height=72mm,bb=30 50 825 525,clip]{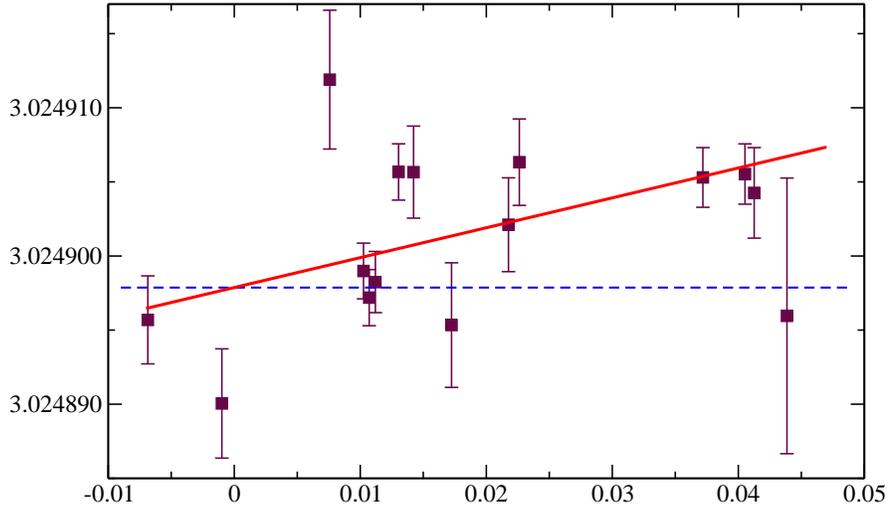}
   \caption{\small Regression analysis of $z_i$-to-$K_i$ for the
                   \h2 lines presented in Table~I.
                   The parameters of the regression line are
   $z_{abs}=3.024898(2)\;$ and $\;b=(20.2\pm7.3)\times10^{-5}$.}
   \end{figure}

\section*{Conclusions}
We obtained that possible deviation of the proton-to-electron mass
ratio at the epoch corresponding about 10~Gyrs ago is the
following for unweighted points $(\Delta \mu / \mu)_{\rm u} =
(3.0\pm2.4)\times10^{-5}$ and for weighted points
\[
(\Delta \mu / \mu)_{\rm w} = (5.02\pm1.82)\times10^{-5}\, .
\]
The significance of the both results is less than $3\sigma$,
therefore, it may be considered only as a glimpse on possible
cosmological variation of $\mu$.

In order to improve the result it is necessary: \\
(i) to measure \h2 laboratory wavelengths with better accuracy; \\
(ii) to refine the calibration procedure; \\
(iii) to study \h2 absorption systems in spectra of other quasars.

In any case, this result may be considered as the most stringent
estimate on an upper limit of a possible variation of $\mu$ (95\%
C.L.):
\[
|\Delta \mu / \mu| < 8\times 10^{-5}\, .
\]

\acknowledgements \theendnotes The observations have been obtained
with UVES mounted on the \mbox{8.2-m} KUEYEN telescope operated by
the European Southern Observatory at Parana, Chili. A.~Ivanchik
and D.~Varshalovich are grateful to the RFBR (02-02-16278-a,
02-02-06096-mac) for support. This work has been supported in part
by scientific exchange program between the RAS (Russia) and the
CNRS (France). A.~Ivanchik is grateful for the opportunity to
visit the IAP-CNRS. P.~Petitjean and E.~Rodriguez thank the Ioffe
Institute for hospitality during the time part of this work was
completed.

\end{article}
\end{document}